\documentclass[%
article,
superscriptaddress, letter, twocolumn,
amsmath,amssymb,
prb,
]{revtex4-1}

\usepackage[english]{babel}
\usepackage[utf8]{inputenc}
\usepackage{ae}
\usepackage{aecompl}
\usepackage{lipsum}
\usepackage{graphicx}
\usepackage{dcolumn}
\usepackage{bm}
 
\usepackage{siunitx}

\begin{document}

\title{Radical-induced Hetero-Nuclear Mixing and Low-field $^{13}$C Relaxation in Solid Pyruvic Acid} 

\author{Hana Kou\v{r}ilov\'{a}}
\email{hana.kourilova@kit.edu}
\author{Michael Jurkutat}
\email{michael.jurkutat@kit.edu}
\affiliation{Institute of Biological Interfaces 4, Karlsruhe Institute of Technology, Germany}
\author{David Peat}
\affiliation{School of Physics and Astronomy, University of Nottingham, Nottingham, NG7 2RD, UK}
\author{Karel Kou\v{r}il}
\affiliation{Institute of Biological Interfaces 4, Karlsruhe Institute of Technology, Germany}
\author{Alixander S. Khan}
\affiliation{School of Physics and Astronomy, University of Nottingham, Nottingham, NG7 2RD, UK}
\author{Anthony J. Horsewill}
\author{James F. MacDonald}
\author{John Owers-Bradley}
\affiliation{School of Physics and Astronomy, University of Nottingham, Nottingham, NG7 2RD, UK}
\author{Benno Meier}
\email{benno.meier@kit.edu}
\affiliation{Institute of Biological Interfaces 4, Karlsruhe Institute of Technology, Germany}
\affiliation{Institute of Physical Chemistry, Karlsruhe Institute of Technology, Germany}
 
\date{\today}

\newcommand{\bH}{\beta_\mathrm{H}}
\newcommand{\bC}{\beta_\mathrm{C}}
\newcommand{\bNZ}{\beta_\mathrm{NZ}}

\newcommand{\tsi}{\tau_\mathrm{SSI}}
\newcommand{\tsc}{\tau_\mathrm{NZ-C}}
\newcommand{\tsh}{\tau_\mathrm{NZ-H}}

\begin{abstract}
Radicals serve as source in dynamic nuclear polarization, but may also act as polarization sink.
If the coupling between the electron spins and different nuclear reservoirs is stronger than any of the reservoirs' couplings to the lattice, radicals can mediate hetero-nuclear mixing.
Here, we report radical-enhanced $^{13}$C relaxation in pyruvic acid doped with trityl.
We find a linear dependence of the carbon $T_1$ on field between 5\,mT and 2\,T.
We extend a model, employed previously for protons, to carbon, and predict efficient proton-carbon mixing via the radical Non-Zeeman reservoir, for fields from 20\,mT to beyond 1\,T.
Discrepancies between the observed carbon relaxation and the model are attributed to enhanced direct hetero-nuclear mixing due to trityl-induced linebroadening, and a field-dependent carbon diffusion from the radical vicinity to the bulk.
Measurements of hetero-nuclear polarization transfer up to 600\,mT confirm the predicted mixing as well as both effects inferred from the relaxation analysis.
 \end{abstract}

\pacs{Valid PACS appear here}
\maketitle

%

\section*{Article}

Nuclear magnetic resonance (NMR) and magnetic resonance imaging (MRI) are powerful tools in various fields of science and industry, but their versatility is limited by the weak nuclear polarization. 
The low sensitivity may be alleviated by dynamic nuclear polarization (DNP), where the large polarization of electron spins is transferred to nuclear spins. 
In dissolution-DNP,\cite{ardenkjaer-larsen-2003-increas-signal} this process is carried out at temperatures near 1 K and fields of several Tesla. 
Under these conditions, the electron spins are almost fully polarized, and it is possible to achieve near unity polarization also for nuclear spins. 
The hyperpolarized sample is then dissolved with a jet of hot solvent, and the solution is used for sensitized NMR spectroscopy\cite{ardenkjaer-larsen-2019-intro,koeckenberger-2014-dissol-dynam,kurzbach-dissol-dynam-2018} or imaging\cite{nelson-2013-metab-imagin,wang-hyper-mri-2019,gallagher-imagin-breas-cancer-2020,woitek21_hyper_carbon_mri_early_respon}.

In bullet-DNP, the order of dissolution and transfer is reversed.\cite{kouril-2019-scalab-dissol, kouril21_cryog_free_semi_autom_appar} 
Here, the solid sample is transferred rapidly to the second magnet, and dissolved only near the NMR tube. 
This procedure limits dilution, avoids the use of hot solvents, and may be beneficial for hyperpolarization of moieties with a short $T_1$ in the liquid state.

During the bullet transfer at low field (currently about 70\,mT) the radical spin that is needed as source for DNP, however, may act as polarization sink. 
The extent of this effect was previously difficult to estimate, since the low-field low-temperature relaxation in presence of radicals has not been studied comprehensively.
A previous study at low temperature in pyruvic acid doped with trityl between 0.9 and 9\,T found a cubic field-dependence of the carbon $T_1$.\cite{niedbalski-2018-magnet-field}
If this trend were to continue to lower fields, it would be detrimental for bullet-DNP.

We have recently reported the low-temperature low-field $^1$H relaxation in pyruvic acid doped with trityl, where we showed that a first-principle based formalism by Wenckebach can describe the effect of trityl quantitatively over two orders of magnitude in field.
Here, we present the corresponding $^{13}$C relaxation measurements as well as thermal mixing data in non-degassed neat 1-$^{13}$C pyruvic acid (neat PA, $^\circ$) and 1-$^{13}$C pyruvic acid doped with 15 mM OX063 (doped PA, $^\bullet$) for fields up to 2\,T.
The measurements were performed using a fast field cycling (FFC) setup previously described\cite{horsewill-2002-magnetic, peat-2016-low-field}, experimental details are given in the \textit{Supplement}. 

\begin{figure}[th!]
  \centering
  \includegraphics[width=0.4\textwidth]{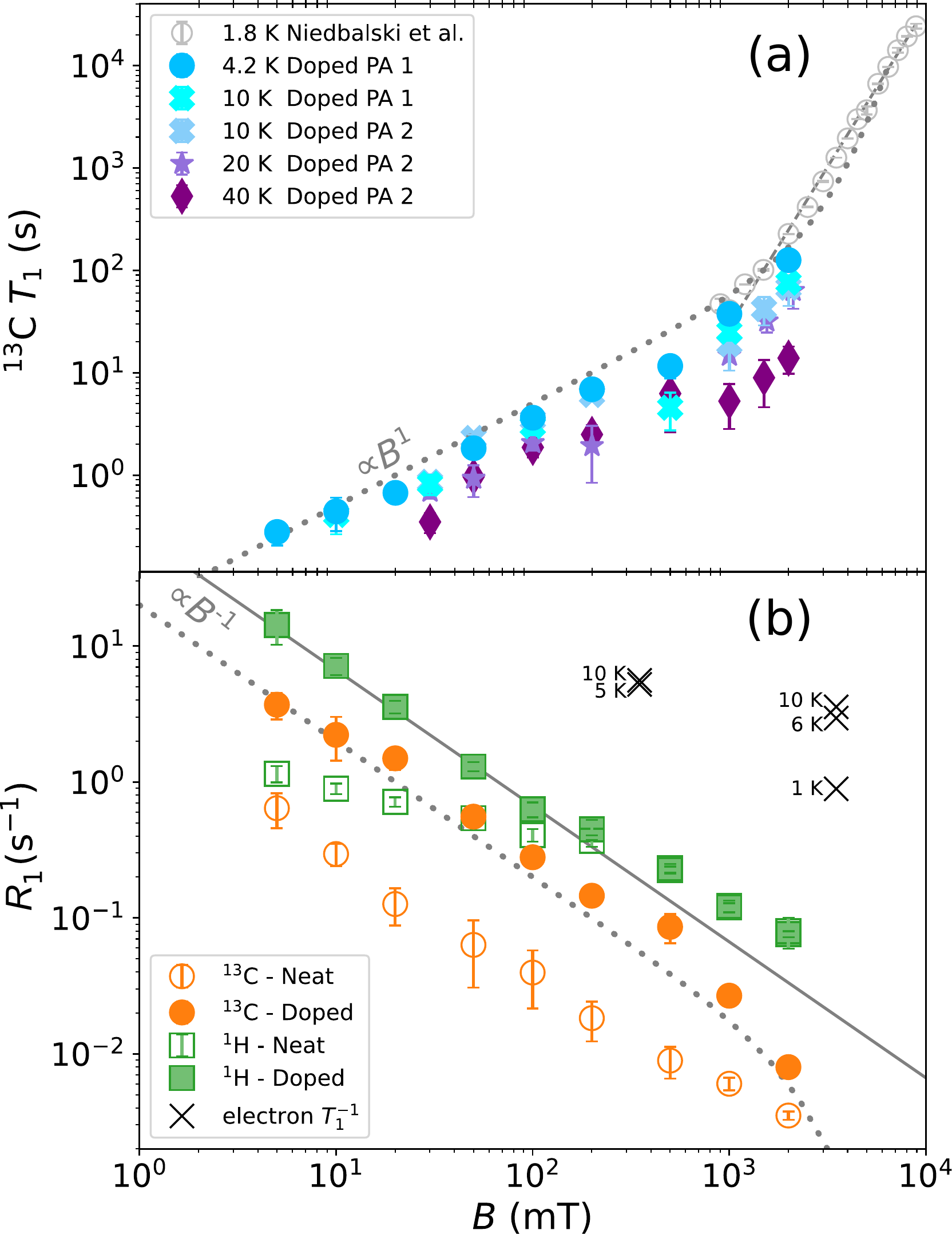}
  \caption{Field-dependent relaxation in 1-$^{13}$C PA doped with 15\,mM trityl (OX063). 
  (a) $^{13}$C relaxation times for different temperatures together with 1.8 K data (gray) reproduced from Ref.~\cite{niedbalski-2018-magnet-field}. The dashed gray line corresponds to a power law with an exponent $\alpha = 3.1$ as reported by Niedbalski et al. based on measurements at 1.8\,K and fields from 0.9 to 9\,T. The dotted gray line is computed using $T_{1} = c B / (1 - P_0^2)$, where $c = 50$ s/T, and $P_0 = \tanh ( \hbar\gamma_{S}B / (2 k_B T))$ is the electron spin polarization. 
  (b) The 4.2~K data on doped PA (solid symbols) plotted as rates together with $^1$H data. For comparison the relaxation rates measured on neat PA (open symbols) are shown as well. The presence of OX063 increases the relaxation rate of carbons throughout the observed field range, and causes faster proton relaxation up to approximately 200\,mT. Also shown are EPR relaxation data reported by Lumata et al.\cite{lumata-2013-elect-spin} near 4.2\,K.}
  \label{fig:T1Data}
\end{figure}

Fig.\,\ref{fig:T1Data}(a) shows the field-dependence of the carbon relaxation time $T_\mathrm{1,C}^\bullet$ for doped PA for temperatures between 4.2\,K and 40\,K.
At low fields, we find an approximately linear increase of $T_\mathrm{1,C}^\bullet$ with field.
This low-field linearity holds up to 40\,K, although $T_\mathrm{1,C}^\bullet$ expectedly decreases with increasing temperature.    

Also shown in Fig.\,\ref{fig:T1Data}(a) are data at 1.8\,K recorded by Niedbalski, Lumata and co-workers \cite{niedbalski-2018-magnet-field}. 
They found that their data are well described by a power law $T_1 = C B^{\alpha}$, with an exponent of $\alpha = 3.1$, indicated by the dashed gray line. 
While we observe a linear rather than a cubic field dependence at lower fields, our data at 4.2\,K are nonetheless in good agreement with the data by Niedbalski et al. at 1.8\,K.

The steep increase in $^{13}$C $T_\mathrm{1,C}^\bullet$, as observed by Niedbalski et al. at 1.8\,K, may be attributed to the substantial electron spin polarization $P_0$ with increasing field, which leads to a suppression of triple-spin flips. 
As shown in Fig.\,\ref{fig:T1Data}(a), a corresponding correction of the observed linear dependence (dotted gray line) gives a satisfactory description over more than four orders of magnitude in $T_1$.

To assess the effect of trityl radicals on nuclear relaxation, it is requisite to compare the relaxation of both $^1$H and $^{13}$C nuclear spins in the presence of trityl to that in neat PA. 
Relaxation rates $R_1=1/T_1$ of both nuclei, recorded at 4.2 K, are shown for neat and doped PA in Figure \ref{fig:T1Data}(b). 

We note for neat PA that the rates $R_\mathrm{1}^\circ$ of protons and carbons converge at low field, which is expected based on reports of direct low-field thermal mixing in neat PA \cite{hirsch-2015-brute-force,peat-2016-low-field}.
This direct exchange is negligible for $B > 10$\,mT but becomes relevant as the field approaches zero.

The presence of trityl significantly accelerates the relaxation of carbon over the entire field range, while the effect on proton relaxation is only observable for fields below 200\,mT. 

In addition to the nuclear relaxation rates, Fig.\,\ref{fig:T1Data}(b) also shows electron spin-lattice relaxation rates reported by Lumata et al. \cite{lumata-2013-elect-spin} for temperatures near 4.2\,K.
For our analysis below, we note that $R_\mathrm{1,S} (T=4 \mathrm{K}) \approx 5\,\mathrm{s}^{-1}$ is nearly independent of field, which is expected since trityl relaxes predominantly via oxygen.\cite{hess12_measur_elect_spin_lattic_relax}

\begin{figure}[t!]
  \centering
  \includegraphics[width=0.425\textwidth]{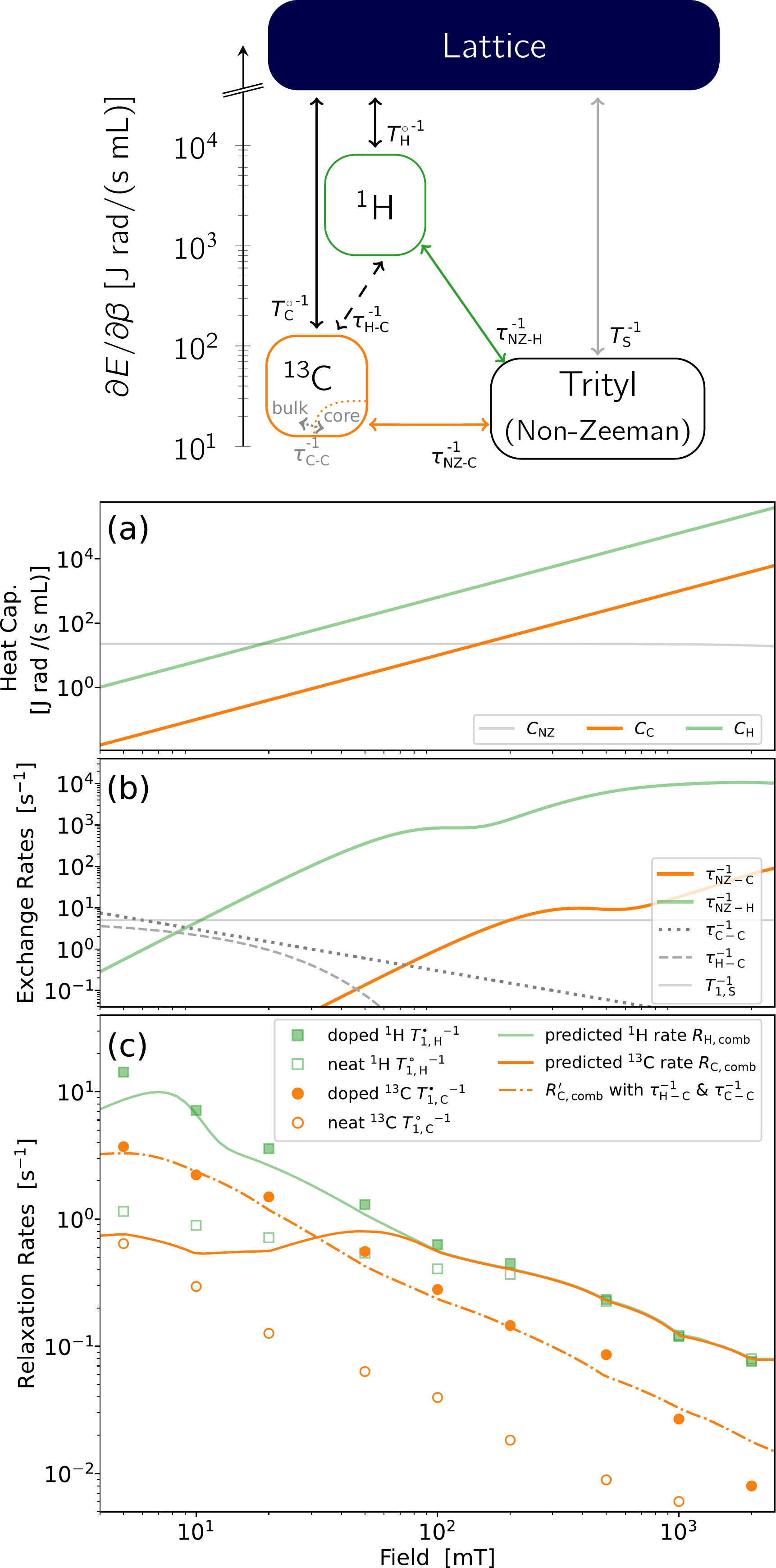}
  \caption{ \textit{top}: Carbon relaxation is described by three reservoirs coupled to the lattice and each other. (heat capacity for 1-$^{13}$C PA at 200\,mT as vertical axis)
  (a) Heat capacities of nuclear Zeeman reservoirs grow quadratically with field, while the NZ heat capacity is field-independent.
  (b) Exchange rates used for modelling of nuclear relaxation rates. The TSF rates $\tau_\mathrm{NZ-H/C}^{-1}$ have previously been calculated\cite{jurkutat-2022} and the electron $T_\mathrm{1,S}^{-1}$ is deduced from literature data\cite{lumata-2013-elect-spin}. Also shown are the deduced carbon core-bulk diffusion rate $\tau_\mathrm{C-C}^{-1}$ and the proton-carbon mixing rate $\tau_\mathrm{H-C}^{-1}$.
  (c) The resulting coefficient-weighted relaxation rates for protons $R_\mathrm{H,comb}(t=T_\mathrm{1,H}^\bullet)$ describe $^1$H data well \cite{jurkutat-2022}. 
  The predicted rate for carbon $R_\mathrm{C,comb}(t=T_\mathrm{1,C}^\bullet)$ is slower than experimental rates ${T_\mathrm{1,C}^\bullet}^{-1}$ for fields up to 30\,mT and faster above. 
  Between 0.1 and 1\,T the high TSF rates for both nuclei cause the predicted rate for carbon to be tied to the proton relaxation.
  Taking into account faster low-field mixing with protons and carbon-carbon spin-diffusion, the $^{13}$C data are well described (dash-dot line).
  } 
  \label{fig:T1calcs}
\end{figure}

We have recently shown that the proton relaxation can be described quantitatively\cite{jurkutat-2022} using a thermodynamic spin temperature model.\cite{abragam_experiments_1957,abragam_spin_1958,Cox_1973}
In this model, as indicated in the sketch in Fig.\,\ref{fig:T1calcs}, the different nuclear spin species and radical electron spins are described as reservoirs with distinct inverse temperatures $\beta_i = \hbar / (k_B T_i)$ that couple to one another and the lattice.
In particular, we showed that the relaxation of proton spins is described by their coupling to the electron Non-Zeeman (NZ) or dipolar reservoir.
This exchange is via energy-conserving triple-spin flips (TSFs), in which a nuclear spin flip and an electron-electron flip-flop occur simultaneously.\cite{wenckebach-2016-essentials}
The TSF rate was calculated from first principles,\cite{wenckebach-2019-dynam-nuclear,wenckebach-2019-dynam-nuclear2} after determining the NZ heat capacity from proton relaxation data. 
This lead to a nearly quantitative agreement between the model and the measured proton relaxation rates over two orders of magnitude in magnetic field.\cite{jurkutat-2022}

Note that in this previous analysis of the proton relaxation, the carbon reservoir could be ignored, as its heat capacity $C_\mathrm{C}$ is about 64 times smaller than that of the proton reservoir $C_\mathrm{H}$ for pyruvic acid at any field, as shown in Fig.\,\ref{fig:T1calcs}\,(a). 
We now extend the analysis to the carbon reservoir, which is coupled to the lattice with the rate ${T_\mathrm{1,C}^\circ}^{-1}$ measured in neat PA and to the NZ reservoir with the TSF rate $\tau_\mathrm{NZ-C}^{-1}$.
The latter is calculated\cite{jurkutat-2022} and compared to the TSF rate for protons in Fig.\,\ref{fig:T1calcs}\,(b).
The field-dependence of the carbon TSF rate is shifted relative to that of protons by a factor of $\gamma_\mathrm{H}/\gamma_\mathrm{C}=4$ and its amplitude is two orders of magnitude smaller.

Details on the extension of the previous analysis\cite{jurkutat-2022} to three reservoirs and its numerical solution can be found in the \textit{Supplement}. 
The relaxation of three coupled reservoirs to the lattice temperature is generally given by a tri-exponential decay. 
Since the experimental data do not warrant an extraction of three coefficients and three decay rates, we compare the experimental data in Fig.\,\ref{fig:T1calcs}\,(c) with effective relaxation rates $R_\mathrm{i,comb}(t)=-{\dot{\beta}_\mathrm{i}}(t)/{{\beta_\mathrm{i}}(t)}$, evaluated at the time of measurement $t=T_{1,i}^\bullet (B_0)$. 
Details are givene in the \textit{Supplement}. 

As expected, the presence of the carbon reservoir does not affect the resulting proton relaxation (solid green line in Fig.\,\ref{fig:T1calcs}\,(c)), i.e., the three-reservoir description traces the proton data for doped PA (full squares) as well as the previous analysis,\cite{jurkutat-2022} which had considered only protons and the NZ reservoir.

The solution for the expected carbon rate (solid orange line in Fig.\,\ref{fig:T1calcs}\,(c)), however, does not describe the measured carbon rates well.
At fields up to 20 mT, the predicted relaxation rates are smaller than the ones observed in experiment.
For fields above 50 mT, the predicted relaxation rates are larger than the ones observed in experiment. 

The discrepancy at low fields may be attributed to direct thermal mixing between the $^1$H and $^{13}$C reservoirs.
The rate for this process, denoted by $\tau_{\rm{H-C}}^{-1}$ in Fig.\,\ref{fig:T1calcs}(b), is negligible above 20\,mT but becomes significant at lower fields.\cite{peat-2016-low-field}
Note that this direct mixing process causes the convergence, noted above, of the experimental carbon and proton relaxation rates at low fields in neat PA.
Therefore, in neat PA below 20\,mT, $\tau_{\rm{H-C}}^{-1}$ should correspond to the experimentally observed carbon relaxation rate. 
We find that the low-field relaxation rates of $^{13}$C in doped PA are well described by  $5 \cdot {T_{\mathrm{1,C}}^{\circ}}^{-1}$, i.e. the direct proton-carbon relaxation rates scale with the neat rates, but are accelerated five-fold.
We attribute this increase to line-broadening due to the presence of trityl in doped PA.

The discrepancy observed at fields above 50\,mT may be attributed to carbon spin diffusion. 
In the model, the protons are in the fast thermal mixing limit in this field range, i.e., the NZ reservoir exchanges faster with the protons than with the lattice ($\tau_\mathrm{NZ-H}^{-1} \gg T_\mathrm{1,S}^{-1}$) and the NZ heat capacity does not affect the proton reservoir relevantly ($C_\mathrm{NZ} \ll C_\mathrm{H}$).  
Therefore the proton reservoir sets the temperature of the Non-Zeeman reservoir.
As the field is increased, the carbon reservoir increasingly couples to the NZ reservoir, and so the model predicts the same spin temperatures and hence relaxation rates for both protons and carbons.

However, only a minute portion of nuclear spins in the radical vicinity, referred to as \textit{core spins}, are in direct exchange with the electron NZ reservoir. 
Carbon spins in the \textit{bulk} only exchange the radicals indirectly via carbon spin diffusion.
From our experiments we cannot distinguish whether a slow exchange between core and bulk carbon spins impedes the relaxation, or if the diffusion through the bulk is field-dependent, and we discuss both possibilities below

As detailed in the \textit{Supplement} and indicated in the sketch in Fig.\,\ref{fig:T1calcs}, the relaxation model may be extended with a second carbon reservoir and a field-dependent carbon core-bulk time constant of 33\,s/T.
This leads to a satisfactory description (dash-dot line in Fig.~\ref{fig:T1calcs}\,(c)) of the experimental data.

So altogether, the model, that describes low-field proton relaxation in trityl-doped PA by exchange with the radical NZ reservoir \cite{jurkutat-2022}, can be expanded to account for the $^{13}$C relaxation.
The model predicts efficient proton-carbon coupling mediated by the NZ reservoir above 20\,mT, and our analysis indicates additionally (i) a five-fold accelerated direct hetero-nuclear exchange at low field, and (ii) that the indirect hetero-nuclear coupling is slowed by a field-dependent carbon diffusion.
  
\begin{figure*}[th!]
  \centering
  \includegraphics[width=0.99\textwidth]{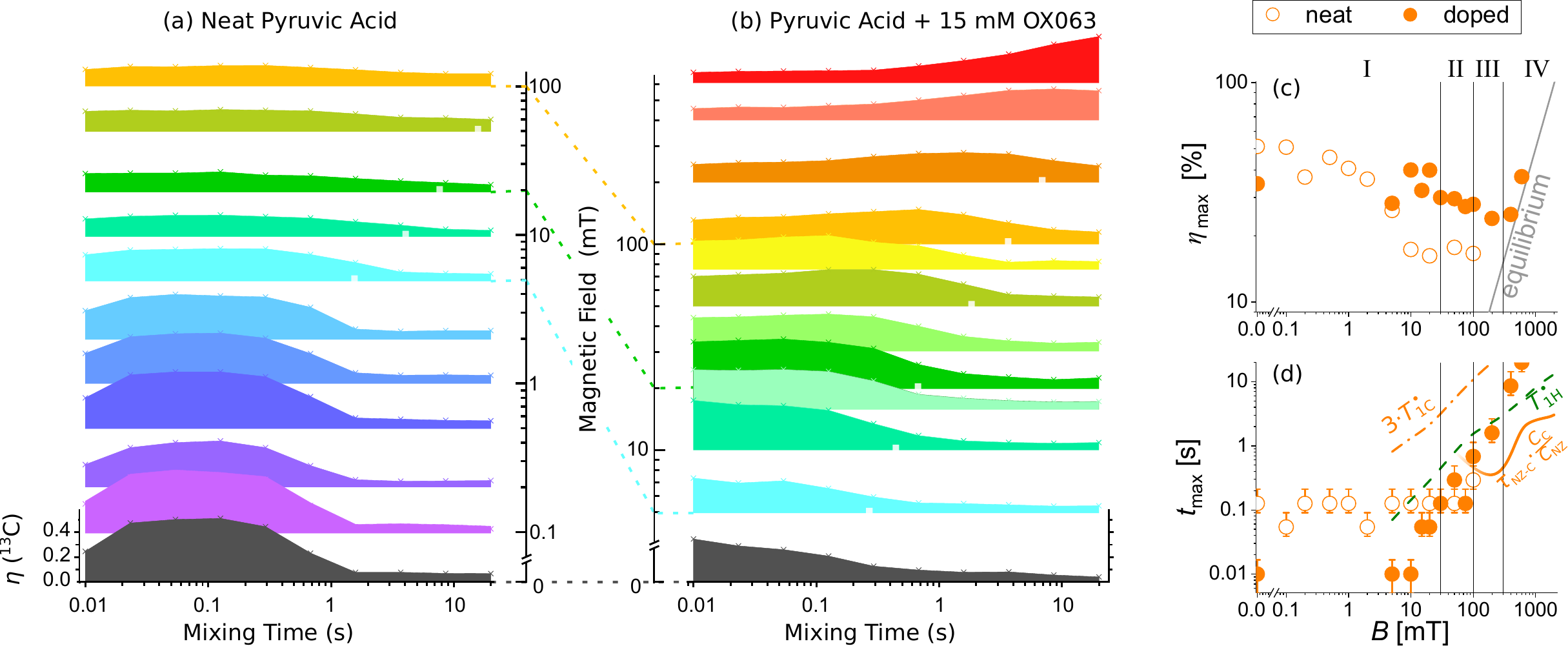}
  \caption{Thermal mixing in (a) neat and (b) doped PA at 4.2\,K for fields from zero up to 200\,mT and 600\,mT, respectively. 
  All data were normalized to the thermal equilibrium $^{13}$C signal at 4.2\,K and 2 T (see supporting information), with the scale indicated for the respective zero-field data (\textit{lower left}). 
TM efficiency is shown as a function of mixing delay. Also the measured $^{13}$C $T_1$  are indicated, where available, by \textit{white squares} in the respective datasets.  
(c) The field-dependent maximum mixing efficiency $\eta_\mathrm{max}$ shows that trityl mediates hetero-nuclear mixing above 5\,mT, since $\eta_\mathrm{max}$ is above equilibrium polarization throughout the investigated field range. 
(d) The corresponding times for TM maxima, $t_\mathrm{max}$, show that trityl accelerates TM below 20\,mT. 
Above 20\,mT $t_\mathrm{max}(B)$ for doped PA confirms TM, $t_\mathrm{max} \ll 3 \cdot T_\mathrm{1,C}^\bullet$(orange dash-dot line). 
Also, for $B \geq 20 \mathrm{mT}$ we find $t_\mathrm{max}$ grows with field, consistent with TM slowed down by diffusion.
Beyond 200\,mT the mixing process becomes slower than the proton relaxation, $t_\mathrm{max}>T_\mathrm{1,H}$ (green dashed line), such that $\eta_\mathrm{max}$ in (c) only slightly above equilibrium can be achieved by TM for 400 and 600\,mT. }
  \label{fig:TM}
\end{figure*}

We now show that model's prediction as well as the modifications (i,ii) are consistent with the low-field thermal mixing (TM) experiments, which are displayed in Fig.\,\ref{fig:TM}. 
In these experiments the carbon spins are saturated while the proton polarization is allowed to achieve thermal equilibrium at 2\,T. 
The field is then ramped to the mixing field for the duration of the mixing delay. 
Subsequently the field is ramped back to 2\,T, where the $^{13}$C spin polarization is read out. 
Further details are given in the \textit{Supplement}.

In order to quantify the effectiveness of the thermal mixing step we define the thermal mixing efficiency $\eta$ as the observed signal divided by the thermal equilibrium signal of $^{13}$C at 2\,T. 
We can neglect the heat load of the carbon reservoir on the protons, so a thermal mixing efficiency of 100\% then implies that the carbon spins attain the spin temperature of the proton spins without any losses, which corresponds to the theoretical maximum attainable in absence of any relaxation.

The dependences of TM efficiency on mixing delay at different fields measured in neat and doped PA at 4.2\,K are shown in panels (a) and (b) of Fig.\,\ref{fig:TM}, respectively. 
In Fig.\,\ref{fig:TM}\,(c) and (d) the field-dependent maximum TM efficiency $\eta_\mathrm{max}$ and corresponding mixing time $t_\mathrm{max}$ are shown, respectively. 

For neat PA we find thermal mixing is most efficient well below 10\,mT, in agreement with a previous study \cite{peat-2016-low-field}, and the converging $T_1^\circ$s for protons and carbons pointed out above. 
We note, however, a limited polarization transfer to carbon in neat PA at fields up to 100\,mT, cf. Fig.\,\ref{fig:TM}\,(c), corresponding to $\eta \approx 15\,\%$, which has similarly been observed in previous FFC experiments \cite{peat-2016-low-field}. 
We infer that this transfer occurs during the field ramp up to the resonance field, and it may be associated with the presence of oxygen in the sample or exchange with quantized rotational states of the methyl group.

For doped PA we observe, as predicted, efficient thermal mixing also for fields above 20\,mT. 
For all fields values of $\eta_\mathrm{max}$ greater than or equal to equilibrium polarization are achieved, cf. Fig.\,\ref{fig:TM}\,(c).
And these transfer efficiency maxima are established at shorter mixing times than the measured spin-lattice relaxation could account for, i.e. $t_\mathrm{max}<3\cdot T_\mathrm{1,c}^\bullet$ (orange dash-dot line in Fig.\,\ref{fig:TM}\,(d)).

At low field up to 5\,mT we find TM maxima similar to those in the neat sample, although $\eta$ decreases faster for longer mixing delays at low field, which is due to the radical-enhanced relaxation in the doped sample. 
The shorter times $t_\mathrm{max}$ to reach the mixing maxima in doped PA are consistent with (i) the five-fold faster direct proton-carbon mixing inferred in the relaxation analysis.

Above 10\,mT the direct proton-carbon exchange diminishes, while according to the relaxation analysis indirect exchange via the NZ reservoir increases.
Here, the increasing $t_\mathrm{max}$ with field is consistent with (ii) the slowing of this process by a field-dependent core-bulk diffusion.

Beyond 200\,mT the indirect proton-carbon mixing becomes slower than the proton relaxation, $t_\mathrm{max} > T_\mathrm{1,H}^\bullet$ (green dashed line in Fig.\,\ref{fig:TM}\,(d)), such that no pronounced maxima at 400 and 600\,mT are observable, but rather $\eta_\mathrm{max}$ values slightly above thermal equilibrium.

The TM data are qualitatively consistent with the results from the model used to describe the carbon relaxation.
They show that the presence of trityl causes the predicted (indirect) mixing for fields above 20\,mT.
Additionally they indicate (i) trityl accelerates (direct) mixing at low field, and (ii) that the (indirect) mixing above 20\,mT slows down with increasing field.

In conclusion, the relaxation data reported here explain the success of our bullet-DNP experiments \cite{kouril-2019-scalab-dissol,kouril21_cryog_free_semi_autom_appar}.
Earlier work at fields above 1\,T indicated a scaling of $^{13}$C relaxation rates with field according to $T_1 \sim B^3$,\cite{niedbalski-2018-magnet-field} and, consequently, very fast relaxation at low fields.
Conversely, our data show that below 1\,T the relaxation times scale only linearly with the applied field.

The extension of the model, that successfully describes proton relaxation, to the carbon spins predicts indirect proton-carbon exchange mediated by the NZ reservoir.
Qualitatively, this is consistent with the observed enhanced carbon relaxation as well as the enhanced mixing observed above 20\,mT.

For a quantitative description of the carbon relaxation with our model at low field, we need to include enhanced direct proton-carbon mixing at low fields. 
This is also consistent with the shortened times $t_\mathrm{max}$ to attain mixing maxima in doped PA, and we attribute the enhanced direct hetero-nuclear mixing to NMR line broadening due to the presence of trityl.

At fields above 50 mT the experimentally observed relaxation rates are substantially smaller than those predicted. 
Here the model predicts efficient indirect coupling via the NZ reservoir such that carbon relaxation rates are equal to those of protons up to 2\,T and beyond.
Since only core carbons in the radical vicinity exchange directly with the NZ reservoir, we attribute this discrepancy to a field-dependent carbon diffusion process.
This is also consistent with the thermal mixing data, where $t_\mathrm{max}$ grows with increasing field.
A diffusion process that slows with increasing field is also compatible with long DNP build-up times at high field \cite{ardenkjaer-larsen-2018-cryog-free}.
There are two different diffusion processes that may each reduce the coupling of carbon bulk spins to the Non-Zeeman reservoir.

One possibility is that this reduction arises from slow carbon spin diffusion through the bulk. 
At high field the carbon line in 1-$^{13}$C PA has a significant contribution from chemical shift anisotropy (CSA),\cite{macholl-2010-trity-birad} and one may expect spin diffusion to limit the overall $^{13}$C relaxation as the field, and thereby the carbon linewidth, is increased \cite{wang-2021-speed-nuclear}. 
However, at the fields investigated in this study, the carbon line is strongly dominated by dipolar interactions, so that CSA cannot account for the observed field-dependence of the carbon $T_1$.

An alternative explanation of the observed slower carbon relaxation rates is that the proximity to the radicals causes a shift in the resonance frequency of the core spins, which in turn impedes diffusion to the \textit{bulk} spins. 
The \textit{diffusion barrier} separates the carbon spins into NMR-invisible core nuclei in the radical vicinity, and visible bulk nuclei outside the barrier. 
The diffusion among the bulk nuclei is fast, and so the carbon relaxation rate is limited by the exchange between core and bulk nuclei. 
As measured recently by Stern et al.\cite{stern21_direc_obser_hyper_break_throug}, the core spins do nonetheless exchange polarization with the bulk spins. 
This exchange was modelled with two coupled reservoirs representing the core and bulk spins, respectively, where the diffusion barrier limits the exchange of Zeeman energy between these reservoirs.\cite{stern21_direc_obser_hyper_break_throug}
The energy flow across the barrier is frequently attributed to dipolar interactions \cite{Genack_1975,Cox_1977,Dementyev_2008,furman21_spin_diffus_spin_lattic_relax}, and it is conceivable that the coupling of the two reservoirs is field-dependent, since more Zeeman energy has to be transferred across the barrier at higher fields. 
However, no experimental studies of this coupling for $^{13}$C have have been presented to date, nor its field-dependence.


\begin{acknowledgments}
We thank Tom Wenckebach for numerous discussions and assistance with the relaxation model. We thank Lloyd Lumata for discussions and for providing the OX063 low-field EPR spectrum and the high-field $^{13}$C $T_1$ data, and David Gadian for discussions and help with experiments. We thank Matthias Ernst and Malcolm Levitt for discussions. This work has been supported by EPSRC (EP/R031959/1) and by the ``Impuls- und Vernetzungsfonds of the Helmholtz-Association'' under grant VH-NG-1432. This project has received funding from the European Research Council (ERC) under the European Union’s Horizon 2020 research and innovation programme (grant agreement No 951459).
\end{acknowledgments}

%


\end{document}